\documentclass[aps,prl,twocolumn,superscriptaddress,showpacs,preprintnumbers,amsmath,amssymb]{revtex4-1}

\usepackage{graphicx} 
\usepackage{dcolumn}  
\usepackage{epstopdf}
\usepackage{color}

\graphicspath{{ps}}

\begin{document}


\preprint{\vbox{ \hbox{   }
    \hbox{Belle Preprint 2015-22}
    \hbox{KEK Preprint 2015-62}
    \hbox{Dec. 2015}
}}

\title{ \quad\\[1.0cm] First Observation of Doubly Cabibbo-Suppressed Decay of a Charmed Baryon: $\Lambda^{+}_{c} \rightarrow p K^{+} \pi^{-}$}

\noaffiliation
\affiliation{Aligarh Muslim University, Aligarh 202002}
\affiliation{University of the Basque Country UPV/EHU, 48080 Bilbao}
\affiliation{Budker Institute of Nuclear Physics SB RAS, Novosibirsk 630090}
\affiliation{Faculty of Mathematics and Physics, Charles University, 121 16 Prague}
\affiliation{Chonnam National University, Kwangju 660-701}
\affiliation{University of Cincinnati, Cincinnati, Ohio 45221}
\affiliation{Deutsches Elektronen--Synchrotron, 22607 Hamburg}
\affiliation{University of Florida, Gainesville, Florida 32611}
\affiliation{Justus-Liebig-Universit\"at Gie\ss{}en, 35392 Gie\ss{}en}
\affiliation{Gifu University, Gifu 501-1193}
\affiliation{II. Physikalisches Institut, Georg-August-Universit\"at G\"ottingen, 37073 G\"ottingen}
\affiliation{SOKENDAI (The Graduate University for Advanced Studies), Hayama 240-0193}
\affiliation{Hanyang University, Seoul 133-791}
\affiliation{University of Hawaii, Honolulu, Hawaii 96822}
\affiliation{High Energy Accelerator Research Organization (KEK), Tsukuba 305-0801}
\affiliation{IKERBASQUE, Basque Foundation for Science, 48013 Bilbao}
\affiliation{Indian Institute of Technology Bhubaneswar, Satya Nagar 751007}
\affiliation{Indian Institute of Technology Guwahati, Assam 781039}
\affiliation{Indian Institute of Technology Madras, Chennai 600036}
\affiliation{Indiana University, Bloomington, Indiana 47408}
\affiliation{Institute of High Energy Physics, Chinese Academy of Sciences, Beijing 100049}
\affiliation{Institute of High Energy Physics, Vienna 1050}
\affiliation{Institute for High Energy Physics, Protvino 142281}
\affiliation{INFN - Sezione di Torino, 10125 Torino}
\affiliation{Institute for Theoretical and Experimental Physics, Moscow 117218}
\affiliation{J. Stefan Institute, 1000 Ljubljana}
\affiliation{Kanagawa University, Yokohama 221-8686}
\affiliation{Institut f\"ur Experimentelle Kernphysik, Karlsruher Institut f\"ur Technologie, 76131 Karlsruhe}
\affiliation{King Abdulaziz City for Science and Technology, Riyadh 11442}
\affiliation{Korea Institute of Science and Technology Information, Daejeon 305-806}
\affiliation{Korea University, Seoul 136-713}
\affiliation{Kyoto University, Kyoto 606-8502}
\affiliation{Kyungpook National University, Daegu 702-701}
\affiliation{\'Ecole Polytechnique F\'ed\'erale de Lausanne (EPFL), Lausanne 1015}
\affiliation{Faculty of Mathematics and Physics, University of Ljubljana, 1000 Ljubljana}
\affiliation{Ludwig Maximilians University, 80539 Munich}
\affiliation{Luther College, Decorah, Iowa 52101}
\affiliation{University of Maribor, 2000 Maribor}
\affiliation{Max-Planck-Institut f\"ur Physik, 80805 M\"unchen}
\affiliation{School of Physics, University of Melbourne, Victoria 3010}
\affiliation{Moscow Physical Engineering Institute, Moscow 115409}
\affiliation{Moscow Institute of Physics and Technology, Moscow Region 141700}
\affiliation{Graduate School of Science, Nagoya University, Nagoya 464-8602}
\affiliation{Kobayashi-Maskawa Institute, Nagoya University, Nagoya 464-8602}
\affiliation{Nara Women's University, Nara 630-8506}
\affiliation{National Central University, Chung-li 32054}
\affiliation{National United University, Miao Li 36003}
\affiliation{Department of Physics, National Taiwan University, Taipei 10617}
\affiliation{H. Niewodniczanski Institute of Nuclear Physics, Krakow 31-342}
\affiliation{Nippon Dental University, Niigata 951-8580}
\affiliation{Niigata University, Niigata 950-2181}
\affiliation{University of Nova Gorica, 5000 Nova Gorica}
\affiliation{Novosibirsk State University, Novosibirsk 630090}
\affiliation{Osaka City University, Osaka 558-8585}
\affiliation{Pacific Northwest National Laboratory, Richland, Washington 99352}
\affiliation{University of Pittsburgh, Pittsburgh, Pennsylvania 15260}
\affiliation{University of Science and Technology of China, Hefei 230026}
\affiliation{Seoul National University, Seoul 151-742}
\affiliation{Showa Pharmaceutical University, Tokyo 194-8543}
\affiliation{Soongsil University, Seoul 156-743}
\affiliation{University of South Carolina, Columbia, South Carolina 29208}
\affiliation{Sungkyunkwan University, Suwon 440-746}
\affiliation{School of Physics, University of Sydney, NSW 2006}
\affiliation{Department of Physics, Faculty of Science, University of Tabuk, Tabuk 71451}
\affiliation{Tata Institute of Fundamental Research, Mumbai 400005}
\affiliation{Excellence Cluster Universe, Technische Universit\"at M\"unchen, 85748 Garching}
\affiliation{Department of Physics, Technische Universit\"at M\"unchen, 85748 Garching}
\affiliation{Toho University, Funabashi 274-8510}
\affiliation{Department of Physics, Tohoku University, Sendai 980-8578}
\affiliation{Earthquake Research Institute, University of Tokyo, Tokyo 113-0032}
\affiliation{Department of Physics, University of Tokyo, Tokyo 113-0033}
\affiliation{Tokyo Institute of Technology, Tokyo 152-8550}
\affiliation{Tokyo Metropolitan University, Tokyo 192-0397}
\affiliation{University of Torino, 10124 Torino}
\affiliation{CNP, Virginia Polytechnic Institute and State University, Blacksburg, Virginia 24061}
\affiliation{Wayne State University, Detroit, Michigan 48202}
\affiliation{Yamagata University, Yamagata 990-8560}
\affiliation{Yonsei University, Seoul 120-749}
 \author{S.~B.~Yang}\affiliation{Seoul National University, Seoul 151-742} 
 \author{K.~Tanida}\affiliation{Seoul National University, Seoul 151-742} 
 \author{B.~H.~Kim}\affiliation{Seoul National University, Seoul 151-742} 
  \author{I.~Adachi}\affiliation{High Energy Accelerator Research Organization (KEK), Tsukuba 305-0801}\affiliation{SOKENDAI (The Graduate University for Advanced Studies), Hayama 240-0193} 
  \author{H.~Aihara}\affiliation{Department of Physics, University of Tokyo, Tokyo 113-0033} 
  \author{D.~M.~Asner}\affiliation{Pacific Northwest National Laboratory, Richland, Washington 99352} 
  \author{V.~Aulchenko}\affiliation{Budker Institute of Nuclear Physics SB RAS, Novosibirsk 630090}\affiliation{Novosibirsk State University, Novosibirsk 630090} 
  \author{V.~Babu}\affiliation{Tata Institute of Fundamental Research, Mumbai 400005} 
  \author{I.~Badhrees}\affiliation{Department of Physics, Faculty of Science, University of Tabuk, Tabuk 71451}\affiliation{King Abdulaziz City for Science and Technology, Riyadh 11442} 
  \author{A.~M.~Bakich}\affiliation{School of Physics, University of Sydney, NSW 2006} 
% ...
\collaboration{The Belle Collaboration}


\begin{abstract}
 We report the first observation of the decay $\Lambda^{+}_{c} \rightarrow p K^{+} \pi^{-}$ using a 980 $\mathrm{fb^{-1}}$ data sample collected by the Belle detector at the KEKB asymmetric-energy $e^{+}e^{-}$ collider. This is the first doubly Cabibbo-suppressed decay of a charmed baryon to be observed. We measure the branching ratio of this decay with respect to its  Cabibbo-favored counterpart to be $\mathcal{B}(\Lambda^{+}_{c} \rightarrow p K^{+} \pi^{-})/\mathcal{B}(\Lambda^{+}_{c} \rightarrow p K^{-} \pi^{+})=(2.35\pm0.27\pm0.21)\times10^{-3}$, where the uncertainties are statistical and systematic, respectively.

\end{abstract}

\pacs{13.30.Eg, 14.20.Lq}

\maketitle

\tighten

{\renewcommand{\thefootnote}{\fnsymbol{footnote}}}
\setcounter{footnote}{0}
 Several doubly Cabibbo-suppressed (DCS) decays of charmed mesons have been observed. Their measured branching ratios with respect to corresponding Cabibbo-favored (CF) decays play an important role in constraining models of the decay of charmed hadrons and in the study of flavor-$SU(3)$ symmetry~\cite{dcs_mesons}. Because of the smaller production cross sections for charmed baryons, DCS decays of charmed baryons have not yet been observed, and only an upper limit, $\mathcal{B}(\Lambda^{+}_{c} \rightarrow p K^{+} \pi^{-})/\mathcal{B}(\Lambda^{+}_{c} \rightarrow p K^{-} \pi^{+})<0.46\%$ with 90$\%$ confidence level, has been reported by the FOCUS Collaboration~\cite{dcs_focus}. Theoretical calculations of DCS decays of charmed baryons have been limited to two-body decay modes~\cite{dcs_theory1, dcs_theory2}.\par
 In this letter, we report the first observation of the DCS decay $\Lambda^{+}_{c} \rightarrow p K^{+} \pi^{-}$ and the measurement of its branching ratio with respect to its counterpart CF decay $\Lambda^{+}_{c} \rightarrow p K^{-} \pi^{+}$~\cite{charge_conjugate}. Unlike charmed meson decays, internal $W$ emission and $W$ exchange are not suppressed for charmed baryon decays. In previous studies of CF or singly Cabibbo-suppressed (SCS) decays of $\Lambda_c^{+}$ and $\Xi_{c}^{0}$, direct evidence of $W$ exchange and internal $W$ emission has been observed~\cite{scs1, scs2, scs3, scs4, scs5}. When we consider that $W$ exchange is prohibited in $\Lambda^{+}_{c} \rightarrow p K^{+} \pi^{-}$ but allowed in $\Lambda^{+}_{c} \rightarrow p K^{-} \pi^{+}$, the contribution of $W$ exchange to $\Lambda^{+}_{c}$ decays can be estimated by comparing the measured branching ratio with the na\"{\i}ve expectation~\cite{dcs_focus}, $\tan^{4}{\theta_{{\mathrm c}}}$ (0.285$\%$), where $\theta_{{\mathrm c}}$ is the Cabibbo mixing angle~\cite{cabibbo_angle} and $\sin{\theta_{{\mathrm c}}}=0.225\pm0.001$~\cite{pdg}. This approach does not take into account effects of flavor-$SU(3)$ symmetry breaking. \par
 We analyze data taken at or near the $\Upsilon(1S)$, $\Upsilon(2S)$, $\Upsilon(3S)$, $\Upsilon(4S)$, and $\Upsilon(5S)$ resonances collected by the Belle detector at the KEKB asymmetric-energy $e^{+}e^{-}$ collider~\cite{KEKB}. The integrated luminosity of the data sample is 980 $\mathrm{fb^{-1}}$. The Belle detector is a large-solid-angle magnetic spectrometer comprising a silicon vertex detector (SVD)~\cite{svd2}, a central drift chamber (CDC), an array of aerogel threshold Cherenkov counters (ACC), a barrel-like arrangement of time-of-flight scintillation counters (TOF), and an electromagnetic calorimeter comprised of CsI(Tl) crystals (ECL) located inside a superconducting solenoid coil that provides a 1.5~T magnetic field. The detector is described in detail elsewhere~\cite{Belle}. The combined particle identification (PID) likelihoods, $\mathcal{L}(h)$ ($h = p$, $K$, or $\pi$), are derived from ACC and TOF measurements and $dE/dx$ measurements in CDC. The discriminant $\mathcal{R}(h|h^{'})$, defined as $\mathcal{L}(h) / (\mathcal{L}(h) + \mathcal{L}(h^{'}))$, is the ratio of likelihoods for $h$ and $h^{'}$ identification. The electron likelihood ratio, $\mathcal{R}(e)$, for $e$ and $h$ identification is derived from ACC, CDC, and ECL measurements~\cite{eid}. We use samples of $e^{+}e^{-}\rightarrow c\bar{c}$ Monte Carlo (MC) events, which are generated with PYTHIA~\cite{pythia} and EvtGen~\cite{evtgen} and propagated by GEANT3~\cite{geant3} to simulate the detector performance, to estimate reconstruction efficiencies and to study backgrounds. \par
 In this analysis, our selection criteria follow mostly those typically used in other charmed hadron studies at Belle (for example, Ref.~\cite{dcs_mesons, scs3, scs5}). However, our final criteria are determined by a figure-of-merit (FoM) study performed using a control sample of the CF decay ($\Lambda^{+}_{c} \rightarrow p K^{-} \pi^{+}$) in real data, together with sidebands to the DCS signal region. We use this blinded study to optimize the FoM, defined as $n_{\rm sig} /\sqrt{n_{\rm sig}+n_{\rm bkg}}$, where $n_{\rm sig}$ is the fitted yield of the control sample multiplied by the presumed ratio of the DCS and CF decays (0.0025), and $n_{\rm bkg}$ is the number of background events from the sideband region in the DCS decay. \par 
 A $\Lambda_{c}^{+}$ candidate is reconstructed from the three charged hadrons, and all charged tracks are required to have a distance of closest approach to the interaction point (DOCA) less than 2.0 $\mathrm{cm}$ and 0.1 $\mathrm{cm}$ in the beam direction ($z$) and in the transverse ($r$-$\phi$) direction, respectively. The number of SVD hits is also required to be at least one, both in the $z$ and $r$-$\phi$ directions, for each of three charged particles. The charged particles are identified by the PID measurements: $\mathcal{R}(p|h)>0.9$ for both $h =\pi$ and $K$ is required for charged protons, $\mathcal{R}(K|p)>0.4$ and $\mathcal{R}(K|\pi)>0.9$ are required for charged kaons, $\mathcal{R}(\pi|p)>0.4$ and $\mathcal{R}(\pi|K)>0.4$ are required for charged pions, and $\mathcal{R}(e)<0.9$ is required for all charged particles. The identification efficiencies of $p$, $K$, and $\pi$ are 75$\%$, 75$\%$, and 95$\%$, respectively, for the typical momentum range of the decays. Probabilities of misidentifying $h$ as $h^{'}$, $P(h \rightarrow h^{'})$, are 8$\%$ ($P(p \rightarrow K)$), 
5$\%$ ($P(p \rightarrow \pi)$), 
11$\%$ ($P(K \rightarrow \pi)$),
2$\%$ ($P(K \rightarrow p)$),
2$\%$ ($P(\pi \rightarrow K)$), and less than 1$\%$ ($P(\pi \rightarrow p)$) for the typical momentum range. To suppress combinatorial backgrounds, especially from $B$ meson decays, we place a requirement on the scaled momentum: $x_{p}>0.53$, where $x_{p}$ is defined as $p^{*}/\sqrt{E_{\rm cm}^{2}/4-M^{2}}$; here, $E_{\rm cm}$ is the total center-of-mass energy, $p^{*}$ is the momentum in the center-of-mass frame, and $M$ is the mass of the $\Lambda_{c}^{+}$ candidate. In addition, the $\chi^{2}$ value from the common vertex fit of the charged tracks must be less than 40.

\begin{figure}[h]
\includegraphics[width=0.5\textwidth]{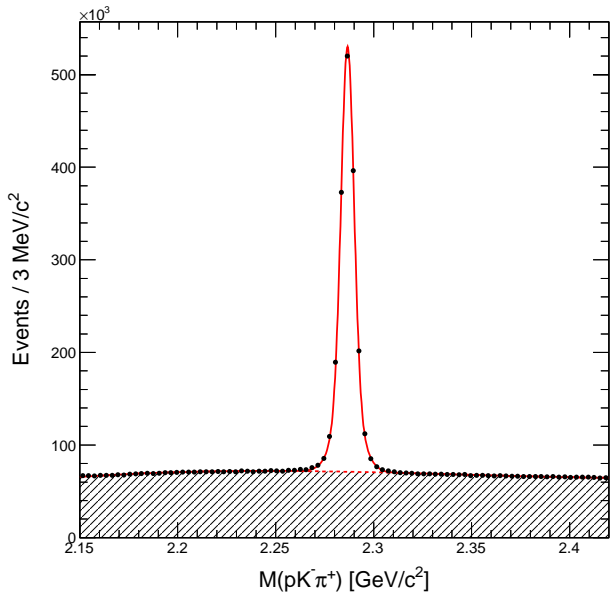}
\caption{Distribution of $M(pK^{-}\pi^{+})$. The curves indicate the fit result: the full fit model (solid) and the combinatoric background only (dashed).
}
\label{fig:cf}
\end{figure}

\begin{figure}[t]
\includegraphics[width=0.5\textwidth]{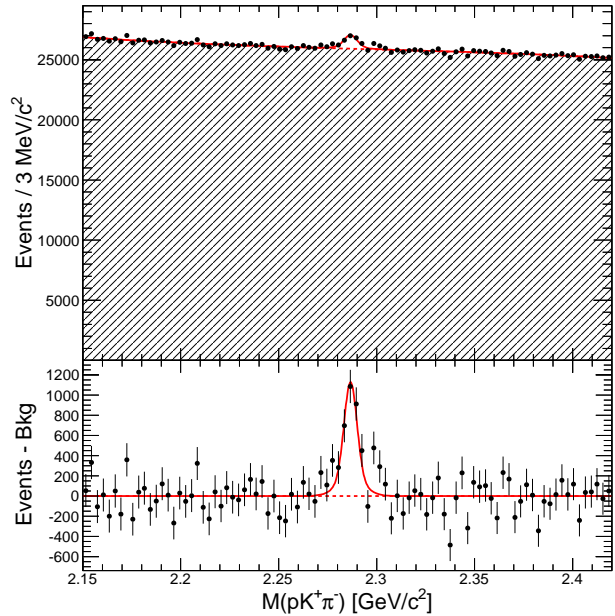}
\caption{Distribution of $M(pK^{+}\pi^{-})$ (top) and residuals of data with respect to the fitted combinatorial background (bottom). Curves are drawn as described in Fig.~\ref{fig:cf}.
}
\label{fig:dcs}
\end{figure}

 Figures~\ref{fig:cf} and \ref{fig:dcs} show invariant mass distributions, $M(pK^{-}\pi^{+})$ (CF) and $M(pK^{+}\pi^{-})$ (DCS), with the final selection criteria. DCS decay events are clearly observed in $M(pK^{+}\pi^{-})$. We perform a binned least-$\chi^{2}$ fit to the two distributions from 2.15~$\mathrm{GeV/}c^{2}$ to 2.42~$\mathrm{GeV/}c^{2}$ with 0.01 $\mathrm{MeV/}c^{2}$ bin width, and the figures are drawn with merged bins. The probability density functions (PDFs) for the fits are the sum of two Gaussian distributions, with a common central value, to represent the signals, and polynomials of fifth and third order for the combinatorial backgrounds in the $M(pK^{-}\pi^{+})$ and $M(pK^{+}\pi^{-})$ distributions, respectively. In the fit to $M(pK^{+}\pi^{-})$, the resolution and central value of the signal function are fixed to be the same as those found from the fit to $M(pK^{-}\pi^{+})$. The reduced $\chi^{2}$ values ($\chi^{2}/d.o.f$) of the fits are 1.03 (27749/26989) and 1.01 (27131/26995) for the CF and DCS decays, respectively. From the fit results, the signal yields of $\Lambda^{+}_{c} \rightarrow p K^{-} \pi^{+}$ and $\Lambda^{+}_{c} \rightarrow p K^{+} \pi^{-}$ decays are determined to be $(1.452\pm0.015)\times10^{6}$ events and $3587\pm380$ events, respectively, where the uncertainties are statistical. There is a small excess above background on the right side of the $\Lambda_{c}^{+}$ peak (around 2.297~$\mathrm{GeV/}c^{2}$) in the DCS spectrum of Fig.~\ref{fig:dcs}. We attribute this to a statistical fluctuation as no known process would make such a narrow feature at this position even when possible particle misidentification, such as the misidentification of both the $K$ and the $\pi$, is taken into account.~\par
 The DCS decay has a peaking background from the SCS decay $\Lambda_{c}^{+}\rightarrow\Lambda K^{+}$ with $\Lambda \rightarrow p\pi^{-}$, which has the same final state topology. However, because of the long $\Lambda $ lifetime, many of the $\Lambda $ vertexes are displaced by several centimeters from the main vertex so the DOCA and $\chi^{2}$ requirements suppress most of this background. The remaining SCS-decay yield is included in the signal yield of $\Lambda^{+}_{c} \rightarrow p K^{+} \pi^{-}$ decay and is estimated via the relation
\begin{multline}
\label{eq:scs_signal_yield}
 \mathcal{N}(SCS;\Lambda \rightarrow p\pi^{-})= \\
\frac {\mathcal{\epsilon} (SCS;\Lambda \rightarrow p\pi^{-})}{\mathcal{\epsilon}(CF)} \frac {\mathcal{B}(SCS;\Lambda \rightarrow p\pi^{-})}{\mathcal{B}(CF)} \mathcal{N}(CF), 
\end{multline}
where $\mathcal{N}(CF)$ is the signal yield of the CF decay, $\mathcal{B}(SCS;\Lambda \rightarrow p\pi^{-})/\mathcal{B}(CF)=(0.61\pm0.13)\%$ is the branching ratio~\cite{pdg}, and $\mathcal{\epsilon} (SCS;\Lambda \rightarrow p\pi^{-})/\mathcal{\epsilon}(CF)=0.023$ is the relative efficiency found using MC samples. This calculation gives a yield of $208\pm78$ events from this source, where the uncertainty is estimated by comparing the signal yields from this calculation and a fit to $M(pK^{+}\pi^{-})$ with loosened selection criteria for the vertex point and $\Lambda $ selection in $M(p\pi^{-})$. After subtraction of this SCS component, the signal yield of the DCS decay is $3379\pm380\pm78$, where the first uncertainty is statistical and the second is systematic due to this subtraction.~\par
 To estimate the statistical significance of the DCS signal, we exclude the SCS signal by vetoing events with $1.1127~\mathrm{GeV/}c^{2}<M(p\pi^{-})<1.1187~\mathrm{GeV/}c^{2}$. The significance is estimated as $\sqrt{-2\ln{(\mathcal{L}_{0}/\mathcal{L})}}$, where $\mathcal{L}_{0}$ and $\mathcal{L}$ are the maximum likelihood values from binned maximum likelihood fits with the signal yield fixed to zero and allowed to float, respectively. The calculated significance corresponds to 9.4$\sigma$. \par
\begin{figure}[htb]
\includegraphics[width=0.5\textwidth, height=0.9\textwidth]{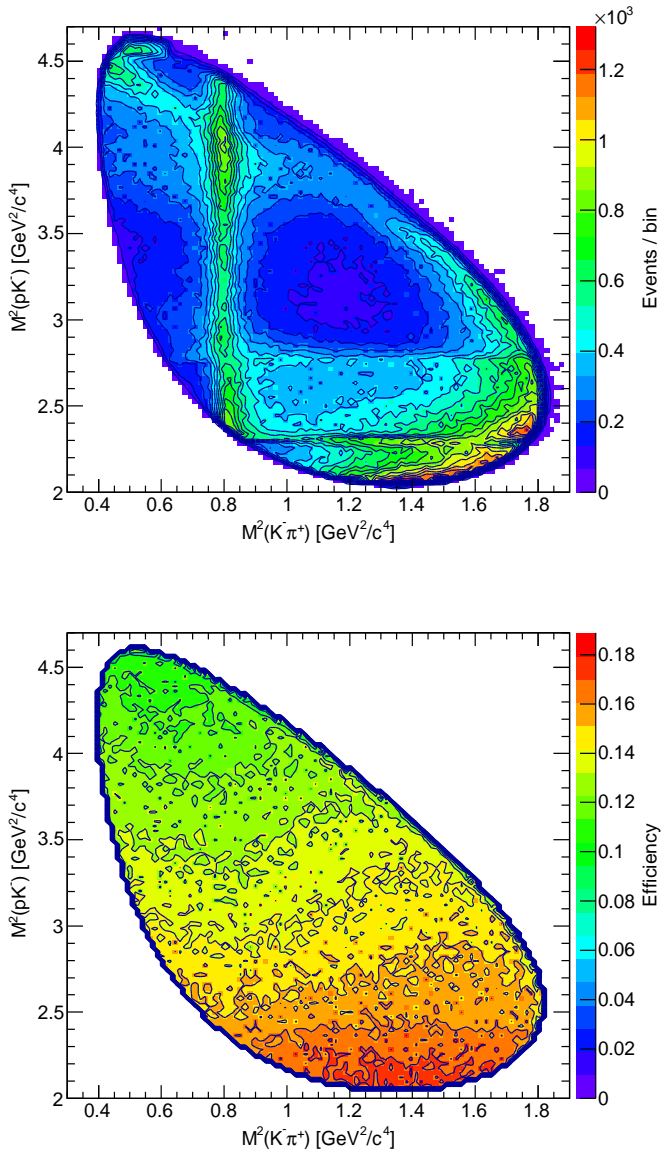}
\caption{Invariant mass squared of $K^{-}\pi^{+}$ versus $pK^{-}$ within $2.2746~\mathrm{GeV/}c^{2}<M(pK^{-}\pi^{+})<2.2986~\mathrm{GeV/}c^{2}$ in real data (top) and estimated efficiency using the MC (bottom). The bin widths of $x$ and $y$ axes are 0.016 $\mathrm{GeV^{2}/}c^{4}$ and 0.027 $\mathrm{GeV^{2}/}c^{4}$, respectively.}
\label{fig:dalitz}
\end{figure}
 We calculate the reconstruction efficiency using a mixture of subchannels weighted with their corresponding branching ratios for the CF decay taken from the PDG~\cite{pdg}. For the DCS decay, we assume subchannels $pK^{*}(892)^{0}$, $\Delta(1232)^{0}K^{+}$, and non-resonant decay with 0.23, 0.18, and 0.59 branching fractions, respectively. These values are chosen as they are the branching fractions for the corresponding subchannels of the CF decay adjusted for the fact that $\Lambda(1520)$ cannot be produced in the DCS decay. To estimate the uncertainty arising from the assumed mix of intermediate states of the CF decay, the reconstruction efficiency is calculated using the efficiency of each bin of the $M^{2}(K^{-}\pi^{+})$ versus $M^{2}(pK^{-})$ Dalitz distribution~\cite{dalitz}, shown in Fig.~\ref{fig:dalitz}, and weighting them by the number of events for the bin in real data. The relative difference between the reconstruction efficiencies, before and after this weighting, is 3.0$\%$. For the DCS decay, the largest relative difference (4.5$\%$) between the efficiency of a subchannel and the overall reconstruction efficiency is taken as the efficiency uncertainty. The relative efficiency between the CF and DCS decays is 1.01$\pm$0.05, where the uncertainty is due to the uncertainty in the composition of the intermediate states.\par
\begin{table}[t]
\centering
\caption{ Systematic uncertainties and sources.}
\label{tbl:sys}
\begin{tabular}
 {@{\hspace{0.3cm}}l@{\hspace{0.3cm}} @{\hspace{0.3cm}}c@{\hspace{0.3cm}}}
\hline \hline
Source & Uncertainty (\%) \\
\hline
Background from SCS signal & $\pm 2.3$ \\
Intermediate states & $\pm 5.4$ \\
Binning and fit range (DCS) & $\pm 5.5$\\
Binning and fit range (CF) & $\pm 0.6$ \\
PDF shape (DCS) & $\pm 2.6$ \\
PDF shape (CF) & $\pm 1.4$ \\
MC statistics & $\pm 0.4$ \\
PID & $\pm 2.2$ \\
Charge-conjugate mode& $\pm 1.8$ \\
\hline
Total & $\pm 9.0$ \\
\hline \hline
\end{tabular}
\end{table}
 The branching ratio, $\mathcal{B}(\Lambda^{+}_{c} \rightarrow p K^{+} \pi^{-})/\mathcal{B}(\Lambda^{+}_{c} \rightarrow p K^{-} \pi^{+})$, is $(2.35\pm0.27\pm0.21)\times 10^{-3}$, where the uncertainties are statistical and systematic, respectively.  Sources of the systematic uncertainty and their values are listed in Table~\ref{tbl:sys}. The uncertainty from the binning and range of the fits is estimated by changing the bin width to 3 $\mathrm{MeV/}c^{2}$ and adjusting the fitted range of the invariant mass distributions. The uncertainty due to the PDF shapes is estimated by changing the order of the polynomial background function, by changing the signal function to the sum of three Gaussian distributions, and by fixing the resolution of the signal function to the MC-derived resolution value. The PID uncertainty is determined by data-MC comparison of several control samples. We treat the relative efficiency difference between charge-conjugate modes as a systematic uncertainty.\par
 The branching fraction of the CF decay, $(6.84\pm0.24^{+0.21}_{-0.27})\times 10^{-2}$, was already well-measured in a previous Belle analysis~\cite{br_cf}. Combining that with our measurement, we determine the absolute branching fraction of the DCS decay to $(1.61\pm0.23^{+0.07}_{-0.08})\times 10^{-4}$, where the first uncertainty is the total uncertainty of the branching ratio and the second is uncertainty of the branching fraction of CF decay. This measured branching ratio corresponds to $(0.82 \pm 0.12)\tan^{4}{\theta_{{\mathrm c}}}$, where the uncertainty is the total.\par
 In conclusion, the first DCS decay of a charmed baryon, $\Lambda^{+}_{c} \rightarrow p K^{+} \pi^{-}$, is observed with statistical significance of 9.4$\sigma$. The branching ratio relative to its counterpart CF decay is $(2.35\pm0.27\pm0.21)\times 10^{-3}$, where the first uncertainty is statistical and the second is systematic. This corresponds to $(0.82\pm0.12)\tan^{4}{\theta_{{\mathrm c}}}$, where the uncertainty is the total uncertainty. Na\"{\i}vely, this would indicate that the $W$ exchange does not make a large contribution to $\Lambda_{c}^{+}$ decays.
\par
 We thank the KEKB group for excellent operation of the accelerator; the KEK cryogenics group for efficient solenoid operations; and the KEK computer group, the NII, and PNNL/EMSL for valuable computing and SINET4 network support. We acknowledge support from MEXT, JSPS and Nagoya's TLPRC (Japan); ARC and DIISR (Australia); FWF (Austria); NSFC and CCEPP (China); MSMT (Czechia); CZF, DFG, and VS (Germany); DST (India); INFN (Italy); MOE, MSIP, NRF, GSDC of KISTI, and BK21Plus (Korea); MNiSW and NCN (Poland); MES and RFAAE (Russia); ARRS (Slovenia); IKERBASQUE and UPV/EHU (Spain); SNSF (Switzerland); NSC and MOE (Taiwan); and DOE and NSF (USA).


%


\begin{thebibliography}{30}
\bibitem{dcs_mesons}
B.R.~Ko {\it et al.} (Belle Collaboration), Phys. Rev. Lett. {\bf 102}, 221802 (2009).

\bibitem{dcs_focus}
J.M.~Link {\it et al.} (FOCUS Collaboration), Phys. Lett. B {\bf 624}, 166 (2005).

\bibitem{dcs_theory1}
K.K.~Sharma and R.C.~Verma, Phys. Rev. D {\bf 55}, 7067 (1997).
\bibitem{dcs_theory2}
T.~Uppal, R.C.~Verma, and M.P.~Khanna, Phys. Rev. D {\bf 49}, 3417 (1994).

\bibitem{charge_conjugate}
Unless stated otherwise, charge-conjugate modes are implied throughout this paper.

\bibitem{scs1}
S.~Henderson {\it et al.} (CLEO Collaboration), Phys. Lett. B {\bf 283}, 161 (1992).
\bibitem{scs2}
P.~Avery {\it et al.} (CLEO Collaboration), Phys. Rev. Lett. {\bf 71}, 2391 (1993).
\bibitem{scs3}
K.~Abe {\it et al.} (Belle Collaboration), Phys. Lett. B {\bf 524}, 33 (2002).
\bibitem{scs4}
B.~Aubert {\it et al.} (\textsl{BABAR} Collaboration), Phys. Rev. Lett. {\bf 95}, 142003 (2005).
\bibitem{scs5}
R.~Chistov {\it et al.} (Belle Collaboration), Phys. Rev. D {\bf 88}, 071103(R) (2013).

\bibitem{cabibbo_angle}
 N. Cabibbo, Phys. Rev. Lett. {\bf 10}, 531 (1963).

\bibitem{pdg}
K. A. Olive {\it et al.} (Particle Data Group), Chin. Phys. C {\bf 38}, 090001 (2014).

\bibitem{KEKB}
S.~Kurokawa and E.~Kikutani, Nucl. Instrum. Methods Phys. Res., Sect. A {\bf 499}, 1 (2003),
and other papers included in this volume; T.Abe {\it et al.}, Prog. Theor. Exp. Phys. {\bf 2013}, 03A001 (2013) and references therein.

\bibitem{svd2} Z.~Natkaniec {\it et al.} (Belle SVD2 Group), Nucl. Instrum. Methods Phys. Res., Sect. A {\bf 560}, 1 (2006);
Y. Ushiroda (Belle SVD2 Group), Nucl. Instrum. Methods Phys. Res., Sect. A {\bf 511}, 6 (2003). 

\bibitem{Belle}
A.~Abashian {\it et al.} (Belle Collaboration), Nucl. Instrum. Methods Phys. Res., Sect. A {\bf 479}, 117 (2002); also see detector section in J.Brodzicka {\it et al.}, Prog. Theor. Exp. Phys. {\bf 2012}, 04D001 (2012). 

\bibitem{eid} H.~Hanagaki {\it et al.}, Nucl. Instrum. Methods Phys. Res., Sect. A {\bf 485}, 490 (2002)

\bibitem{pythia}
T. Sj\"{o}strand, S. Mrenna, and P. Skands, J. High Energy Phys. {\bf 05}, 026 (2006).

\bibitem{evtgen}
D. Lange, Nucl. Instrum. Methods Phys. Res., Sect. A {\bf 462}, 152 (2001); T. Sj\"{o}strand, P. Ed\'{e}n, C. Friberg, L. L\"{o}nnblad, G. Miu, S. Mrenna, and E. Norrbin, Comput. Phys. Commun. {\bf 135}, 238 (2001).

\bibitem{geant3}
R. Brun {\it et al.}, GEANT 3.21, CERN Report DD/EE/84-1, 1984.
\bibitem{dalitz}
R.H. Dalitz, Philos. Mag. {\bf 44}, 1068 (1953).

\bibitem{br_cf}
 A.~Zupanc {\it et al.} (Belle Collaboration), Phys. Rev. Lett. {\bf 113}, 042002 (2014).






\end{thebibliography}
\end{document}